\documentclass[twocolumn,aps,prl,showpacs,superscriptaddress,amsfonts,amssymb,amsmath]{revtex4}
\usepackage{graphicx}

\begin{document}

{\bf Jiang {\sl et al} Reply}: In the comment by Konig, Gefen, and
Silva\cite{ref1}, their main point is to question the
approximation we used in solving the Green's function. They believe
that this approximative method is inappropriate to describe
spin-flip-related dephasing processes caused by intradot
interaction, so they believe that our conclusions
are ill-founded because of that. We agree that we indeed used an
approximation in calculating the Green's function. However, we
believe our approximation to be reasonable and the main
conclusion, namely, an intradot electron-electron interaction can
not induce dephasing, should hold.\cite{ref2} In particular,
we emphasize that we have taken the higher-order terms in solving
the Green's function than appeared in their previous
publications.\cite{ref3,ref4} We will compare
the approximation used their papers\cite{ref3,ref4} with that in our
work.\cite{ref2}

Before we make detailed comparisons, we would first make
several remarks:

(i) They and we used the same formulaes to calculate the current.
Those formulaes were advanced by Meir and Wingreen.\cite{ref5}

(ii) In their device as well as ours, the interaction only exists
in the quantum dot (QD) and the other parts of the device should
be non-interacting, so all Green functions (except
$G^<_{dd}(\omega)$) can be exactly expressed using the intradot
Green function $G^r_{dd}(\omega)$. This means if
$G^r_{dd}(\omega)$ is obtained, then all other Green functions as
well the current can be calculated straightforwardly without any
further approximations.

Here we give an example to exactly express $G^r_{sk,d}(\omega)$
and $G^<_{sk,d}(\omega)$ by $G^r_{dd}(\omega)$ in our system.
$G^r_{sk,d}(\omega)$ and $G^<_{sk,d}(\omega)$ are the Fourier
transforms of $G^r_{sk,d}(t)$ and $G^<_{sk,d}(t)$ with
$G^r_{sk,d}(t) \equiv -i\theta(t)<\{c_{sk}(t),
d^{\dagger}_{d}(0)\}>$ and $G^<_{sk,d}(t) \equiv i
<d^{\dagger}_{d} (0) c_{sk}(t)>$. Following the process of Phys.
Rev. B {\bf 50}, 5528  by Jauho et. al,\cite{ref6} we have:
\begin{eqnarray}
 G^r_{sk,d}(\omega) & =& g^r_{sk} t_{s1} G^r_{1d}
  + g^r_{sk} t_{s2} G^r_{2d} \nonumber \\
  & = & ... \nonumber \\
  & = &
\tilde{g}^r_{sk,1} t_{1} G^r_{dd}
  + \tilde{g}^r_{sk,4} t_{4} G^r_{dd} \nonumber
\end{eqnarray}
\begin{eqnarray}
 G^<_{sk,d}(\omega) & =&
\tilde{g}^r_{sk,1} t_{1} G^<_{dd}
  + \tilde{g}^<_{sk,1} t_{1} G^a_{dd} \nonumber \\
  & & +
 \tilde{g}^r_{sk,4} t_{4} G^<_{dd}
  + \tilde{g}^<_{sk,4} t_{4} G^a_{dd} \nonumber
\end{eqnarray}
where $\tilde{g}^{r,<}$ are the Green functions of the device
decoupled to the QD (i.e. with $t_1 =t_4 =0$) and they
can be solved exactly. Similarly, all other Green functions can
also be expressed using $G^r_{dd}(\omega)$ and $G^<_{dd}(\omega)$.
Moreover although the Keldysh Green function $G^<_{dd}(\omega)$
in general can not be expressed by $G^r_{dd}(\omega)$, $\int
d\omega G^<_{dd}(\omega) $, that is actually needed in the
calculating current, can be expressed by $G^r_{dd}(\omega)$. For
example, in their paper,\cite{ref4} they get this relation by
using $I_L+I_R =0$ [see their Eq.(3.8) and (3.11)]. In our
paper,\cite{ref2} we can obtain the corresponding relation at large
$\Gamma$ case by using the steady state condition.

(iii) By using those exact relations among the Green functions, the
current can be expressed solely by the intradot Green function
$G^r_{dd}(\omega)$. In their paper,\cite{ref4} they give those
expressions [see their Eq.(3.9) and (3.12)] as:
\begin{eqnarray}
 I^{(0)}_R &= &
 -\frac{4e}{h} \frac{\Gamma_L \Gamma_R}{\Gamma_L +\Gamma_R}
  \int d\omega Im G^{r(0)}_{dd} (f_L-f_R)
  \nonumber\\
 I^{(1)}_R & = &
  \frac{4e}{h} \sqrt{\Gamma_L\Gamma_R} |t_{ref}|\cos \varphi
  \int d\omega Re G^{r(0)}_{dd}(f_L - f_R)
 \nonumber
\end{eqnarray}
where $I^{(0)}_R$ and $I^{(1)}_R$ are the zeroth-order
($t_{ref}^0$) term and the first-order flux-dependent
($t_{ref}^1$) term, respectively. $G^{r(0)}_{dd}$ is the
zeroth-order term of $G^r_{dd}$, i.e. $G^r_{dd} = G_{dd}^{r(0)} +
G_{dd}^{r(1)}t_{ref} + G_{dd}^{r(2)}t_{ref}^2 +...$. While
$t_{ref}$ is very small, $G^r_{dd}$ is almost same with
$G^{r(0)}_{dd}$. In our work, we use different approach. We
calculate $G^r_{dd}$, then other Green functions, and at last the
current. We emphasize that those two approaches are essentially
the same.

Now we compare the approximations used in their work and ours in
calculating the Green function $G^{r(0)}_{dd}$, or $G^{r}_{dd}$
for a very small $t_{ref}$.

In our work,\cite{ref2} we first exactly calculate the isolated QD
Green functions $g_{dd}^r(\omega)=
[\omega-\epsilon_{d\sigma}-U+Un_{\bar{\sigma}}]/
[(\omega-\epsilon_{d\sigma})(\omega-\epsilon_{d\sigma}-U)]$.
As a second step, we use the Dyson equation to obtain $G^r_{dd}$:
$G^r_{dd} =g^r_{dd} + g^r_{dd} t_1 {\tilde g}^r_{11} t_1 G^r_{dd}
 + g^r_{dd} t_4 {\tilde g}^r_{44} t_4 G^r_{dd}
 + g^r_{dd} t_1 {\tilde g}^r_{14} t_4 G^r_{dd}
 + g^r_{dd} t_4 {\tilde g}^r_{41} t_1 G^r_{dd}$.
We agree that the second step is not exact, but it is a fairly good
approximation while the QD is weakly coupled to other parts of
the system.

Furthermore, if only for analytical results (i) and (ii) in the
first paragraph of the right column on page 3, not for the
numerical calculations, we may loosen the above approximation. We
may first consider that the Green functions ${\tilde G}^r$ for the
system decoupled to the source and drain leads has been exactly
solved. Secondly, using the Dyson equation to obtain the Green
function of the whole system. Then the results (i) and (ii) on
page 3 can also be obtained in a straightforward manner although
${\tilde G}^r$ is still unknown.

Next, let us examine what approximations are used in their
papers.\cite{ref3,ref4} (i) For $U=0$, they get
$G^{r(0)}_{dd}(\omega) = 1/(\omega-\epsilon_d +i0^+)$, where
$\epsilon_d$ is the intradot level. (ii) For $U=\infty$, in
calculating $I^{(1)}$ they take $G^{r(0)}_{dd} =(P_0
+P_{\sigma})/(\omega-\epsilon_d +i0^+) =
\frac{1}{1+f(\epsilon_d)}\frac{1}{\omega-\epsilon_d +i0^+}$.
(iii) For $U=\infty$, in calculating $I^{(0)}$ [i.e. to obtain
Eq.(3.16) from Eq.(3.9) in Ref.[\onlinecite{ref4}] ] they take
$G^{r(0)}_{dd} =(P_0
+P_{\uparrow}+P_{\downarrow})\frac{1}{\omega-\epsilon_d
+i(\Gamma_L+\Gamma_R)/2} = \frac{1}{\omega-\epsilon_d
+i(\Gamma_L+\Gamma_R)/2}$. So they do not calculate
$G^{r(0)}_{dd}$ at all and they directly write down $G^{r(0)}_{dd}$ from
their intuitive picture. In particular, for $U=\infty$ they use different
expressions of $G^{r(0)}_{dd}$ in the currents $I^{(0)}$
and $I^{(1)}$. This is a serious error because there is only
one $G^{r(0)}_{dd}$ and it can not be given two different values.


At last, we reply their other three comments. (1) They comment that $r_T
> 1$ near resonance in our Fig.2 invalidates it as a good
measure of coherence. In fact, this has been emphasized in our
Letter,\cite{ref2} e.g. see the paragraph after Eq.(1), or the
left column of page 3, etc. We mention it here again: If only the
first-order tunneling process exists, $r_T$ describes the degree
of coherence; otherwise when the higher-order tunneling processes
are not negligible , $r_T$ as well as the amplitude of conductance
$G_1$ does not reflect the degree of coherence. In our Letter we
design a system (i.e. open multi-terminal AB setup) in which the
first-order tunneling process dominates, and we carry out a study
of $r_T$ in such a system. (2) They comment that our Eq.(4) is
wrong, as it relies on the single-particle formalism. Notice in
Eq.(4) we discuss the case of $U=0$ [see the paragraph before
Eq.(4)]. (3). They comment that $\Delta G(\phi)$ should be zero at
$\phi =0$. We made an print error in figure capture, $\Delta G$
should be defined as $\Delta G \equiv G(\phi)-G_0$. Here we also
show the curves (see Fig.1 in this reply) for $\Delta G \equiv
G(\phi)-G(\phi =0)$. We gratefully acknowledge them for pointing
this out.

In conclusion, all our results should hold. The e-e interaction
does not induce any dephasing effect and the asymmetric amplitude
does not associate with the dephasing effect.

This work was financed through Grant No.90303016 of NSFC and
US-DOE.\\

Zhao-tan Jiang$^1$, Qing-feng Sun$^{1,\ast}$, X. C. Xie$^{2}$, and
Yupeng Wang$^{1}$\\
$^1$Institute of Physics, Chinese Academy of Sciences, Beijing
100080, China\\
$^2$Department of Physics, Oklahoma State University, Stillwater,
Oklahoma 74078\\

\begin{figure}[ht]
\begin{center}
\includegraphics[height=4.0cm,width=6cm]{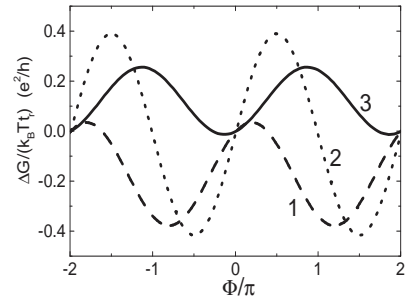}
\end{center}
\caption{ $\Delta G \equiv G(\phi) - G(\phi =0)$ vs $\phi$ with
the same parameters as the Fig.2b in Ref.[\onlinecite{ref2} ] }
\end{figure}


*Electronic mail: sunqf@aphy.iphy.ac.cn

\end{document}